# Mining Spatial Gene Expression Data Using Negative Association Rules

*M.Anandhavalli, M.K.Ghose
Department of Computer Science Engineering
SMIT
Majitar, India

K.Gauthaman
Department of Drug Technology
Higher Institute of Medical Technology
*Derna, Libya*

*Abstract*— Over the years, data mining has attracted most of the attention from the research community. The researchers attempt to develop faster, more scalable algorithms to navigate over the ever increasing volumes of spatial gene expression data in search of meaningful patterns. Association rules are a data mining technique that tries to identify intrinsic patterns in spatial gene expression data. It has been widely used in different applications, a lot of algorithms introduced to discover these rules. However Priori-like algorithms has been used to find positive association rules. In contrast to positive rules, negative rules encapsulate relationship between the occurrences of one set of items with absence of the other set of items. In this paper, an algorithm for mining negative association rules from spatial gene expression data is introduced. The algorithm intends to discover the negative association rules which are complementary to the association rules often generated by Priori like algorithm. Our study shows that negative association rules can be discovered efficiently from spatial gene expression data.

*Keywords- Spatial Gene expression data; Association Rule; Negative Association Rule*

## I.    INTRODUCTION (HEADING 1)

The main contribution here has been a great explosion of genomic data in recent years. This is due to the advances in various high-throughput biotechnologies such as spatial gene expression database. These large genomic data sets are information-rich and often contain much more information than the researchers who generated the data might have anticipated. Such an enormous data volume enables new types of analyses, but also makes it difficult to answer research questions using traditional methods. Analysis of these massive genomic data has two important goals:

1)  To determine how the expression of any particular gene might affect the expression of other genes
2)  To determine what genes are expressed and not expressed as a result of certain cellular conditions, e.g. what genes are expressed in diseased cells that are not expressed in healthy cells?

The most popular pattern discovery method in data mining is association rule mining. Association rule mining was introduced by [4]. It aims to extract interesting correlations, frequent patterns, associations or casual structures among sets of items in transaction databases or other data repositories. The relationships are not based on inherent properties of the data themselves but rather based on the co-occurrence of the items within the database. The associations between items are commonly expressed in the form of association rules. In general, an association rule represents a relationship between two sets of items in the same database. It can be written in the form A → C, where A and C are item sets and A∩C=Φ. The left-hand side (LHS) of the rule is called the antecedent, while the right-hand (RHS) is called consequent. Negative association rules are complementary to the sorts of association rules and have the forms A→¬C or ¬A→C. The rule in the form of ¬A →¬C is equivalent to a positive association rule in the form of C →A.

In this paper, an attempt has been made to study a novel algorithm for discovering of positive association rules and generating meaningful negative association rules in effective manner from spatial gene expression data.

## II.   NOTATIONS

→   means "implies"      U means Union     ¬ means negation

## III.   MATERIALS AND METHODS

### A.   Spatial Gene Expression Data

The Edinburgh Mouse Atlas gene expression database (EMAGE) is being developed as part of the Mouse Gene Expression Information Resource (MGEIR) [1] in collaboration with the Jackson Laboratory, USA. EMAGE (http: //genex.hgu. mrc.ac.uk/Emage/database) is a freely available, curated database of gene expression patterns generated by in situ techniques in the developing mouse embryo [9, 10]. The spatial gene expression data are presented as N×N similarity matrix. Each element in the matrix is a measure of similarity between the corresponding probe pattern and gene-expression region. The similarity is calculated as a fraction of overlap between the two and the total of both areas of the images. This measurement is intuitive, and commonly referred to as the Jaccard index [2, 6]. When a pattern is compared to itself, the Jaccard value is 1 because the two input spatial regions are identical. When it is compared to another pattern, the Jaccard Index will be less than one. If the Jaccard Index is 0, the two patterns do not intersect. If a Jaccard Index value is close to 1, then the two patterns are more similar.

However, biologists are more interested in how gene expression changes under different probe patterns. Thus, these similarity values are discretized such that similarity measure greater than some predetermined thresholds and converted into Boolean matrix.

*B. Data Preprocessing*

Preprocessing is often required before applying any data mining algorithms to improve performance of the results. The preprocessing procedures are used to scale the data value either 0 or 1. The values contained in the spatial gene expression matrix had to be transformed into Boolean values by a so-called discretization phase. In our context, each quantitative value has given rise to the effect of discretization procedure [2]: Max minus x% method.

Max minus x% procedure [7] consists of identifying the highest expression value (HV) in the data matrix, and defining a value of 1 for the expression of the gene in the given data when the expression value was above HV – x% of HV where x is an integer value. Otherwise, the expression of the gene was assigned a value of 0 (Figure 1).

In the similarity matrix, the items I are genes from the data set, where a transaction T⊂ I consists of genes that all have an expression pattern intersecting with the same probe pattern.

The sets of transactions are constructed by taking, for each probe pattern r, every gene g from which its associated gene expression pattern ge satisfies the minimum similarity β, i.e., similarity(r, ge) > β, to form the itemsets.

|   | α (Input) | α (after discretization) |
|---|---|---|
| a | 0.096595 | 0 |
| b | 0.123447 | 0 |
| c | 0.291310 | 1 |
| d | 0.126024 | 0 |
| e | 0.155819 | 0 |
| f | 0.288394 | 1 |
| g | 0.000000 | 0 |
| h | 0.215049 | 1 |

Figure 1. Results of Max minus 25% discretization method

*C. Association Rule Mining*

The Apriori-like algorithms adopt an iterative method to discover frequent itemsets. The process of discovering frequent itemsets need multiple passes over the data. .The algorithm starts from frequent 1-itemsets until all maximum frequent itemsets are discovered. The Apriori-like algorithms consist of two major procedures: the join procedure and the prune procedure. The join procedure combines two frequent k-itemsets, which have the same (k-1)-prefix, to generate a (k+1)-itemset as a new preliminary candidate. Following the join procedure, the prune procedure is used to remove from the preliminary candidate set all itemsets whose k-subset is not a frequent itemsets [3].

From every frequent itemset of k>=2, two subsets A and C, are constructed in such a way that one subset C, contains exactly one item in it and remaining k-1 items will go to the other subset A. By the downward closure properties of the frequent itemsets these two subsets are also frequent and their support is already calculated. Now these two subsets may generate a rule A →C, if the confidence of the rule is greater than or equal to the specified minimum confidence.

*D. Algorithm Details*

1. Let I={i1, i2, …, in} be a set of items, where each item ij corresponds to a value of an attribute and is a member of some attribute domain Dh={d1, d2, …, ds}, i.e. ij Є Dh. If I is a binary attribute, then the Dom (I)={0,1}.

2. A transaction database is a database containing transactions T in the form of (d, E), where d Є Dom(D) and E Є I.

3. Let D be a transaction database, n be the number of transactions in D, and minsup be the minimum support of D. The new_support is defined as new_support = minsup × n.

4. Proposition 1: According to [8], By Boolean vector with AND operation, if the sum of '1' in a row vector Bi is smaller than k, it is not necessary for Bi to involve in the calculation of the k- supports.

5. Proposition 2: According to [5], Suppose Itemsets X is a k-itemsets; $|F_{K-1}(j)|$ presents the number of items 'j' in the frequent set $F_{K-1}$. There is an item j in X. If $|F_{K-1}(j)|$ is smaller than k-1, itemset X is not a frequent itemsets .

6. Proposition 3: $|F_K|$ presents the number of k-itemsets in the frequent set $F_K$. If $|F_K|$ is smaller than k+1, the maximum length frequent itemsets is k.

7. A positive association rule represents a relationship between two sets of items in the form of A →C, where A⊂ I, C ⊂ I and A∩C=Φ.

8. A positive association rule represents a relationship between two sets of items in the form of A → ¬C or ¬A→C, where A ⊂ I, C ⊂ I and A∩C=Φ.

9. The rule A → ¬C has support s% in the data sets, if s% of transactions in T contain itemset A while do not contain item set C. The support of negative association rule supp(A → ¬C, is the frequency of occurrence of transactions with item set A in the absence of item set C.

10. Let X be the set of transactions that contain all items in A. The rule A →¬C holds in given data set with confidence c%, if c% of transactions in X do not contain item set C. Confidence of negative association rule, conf(A → ¬C), can be calculated with supp(AU¬C)/supp(A).

11. Given supp(A →C) and conf(A →C), the support and confidence of the negative rule A→¬C can be computed as follows:
    supp(A →¬C) = supp(A) – supp(A →C)…….….(1)
    conf(A →¬C) = 1 – conf(A →C) ……………........(2)

12. Given supp(A→C) and conf(A→C), the support and confidence of the negative rule ¬A→C can be computed as follows:
    supp(¬A→C) = supp(C) – supp(C →A)………….(3)

$$\text{conf}(\neg A \rightarrow C) = (\text{supp}(C)/1-\text{supp}(A))\ (1-\text{conf}(C \rightarrow A))\ \ldots(4)$$

The introduced algorithm for finding both RHS negative and LHS negative rules in terms of spatial gene expression data in the form of similarity matrix consists of four phases as follows:

1. Transforming the similarity matrix into the Boolean matrix
2. Generating the set of frequent itemsets using Fast Mining algorithm for spatial gene expression data
3. Generating positive rules using Apriori algorithm.
4. Generating negative rules based on existing positive rules.

A detailed description of the introduced algorithm is described as follows:

Part 1: Algorithm for generating frequent itemsets and positive rules.
Input: Spatial Gene Expression data in similarity matrix (M), the minimum support.
Output: Set of frequent itemsets F.
1. Normalize the data matrix M and transformed into Boolean Matrix B;
   // *Frequent 1-itemset generation*
2. for each column $C_i$ of B
3.   If sum($C_i$) >= new_support
4.     $F_1$ = { $I_i$ };
5.   Else delete $C_i$ from B;
   // *By Proposition 1*
6. for each row $R_j$ of B
7.   If sum($R_j$) < 2
8.     Delete $R_j$ from B;
   // *By Proposition 2 and 3*
9. for (k=2; | $F_{k-1}$| > k-1; k++)
10. {
    // *Join procedure*
11.   Produce k-vectors combination for all columns of B;
12.   for each k-vectors combination { $B_{i1}$, $B_{i2}$,…$B_{ik}$ }
13.     { E = $B_{i1} \cap B_{i2} \cap \ldots \cap B_{ik}$
14.       If sum(E) >= new_support
15.       $F_k$ = { $I_{i1}$, $I_{i2}$,…$I_{ik}$ }
16.     }
    // *Prune procedure*
17.   for each item $I_i$ in $F_k$
18.     If |$F_k(I_i)$| < k
19.       Delete the column $B_i$ according to item $I_i$ from B;
20.   for each row $R_j$ of B
21.     If sum($B_j$) < k+1
22.       Delete $B_j$ from B;
23.   k=k+1
24. }
25. Return F = $F_1 \cup F_2 \cup \ldots \cup F_k$

Part 2: Algorithm for generating positive and negative association rules.
Input: Set of Frequent (F), minimum support and minimum confidence.
Output: Set of Positive and Negative Association rules.
26. postiveRule = genRule(FreqSet$_k$);
26. Rule = postiveRule;
    // *Generate Negative Rules*
27. for all rules r Є postiveRule.
28.   negativeRuleSets = genNegCand(r);
29.   for all rules tr Є negaitveRuleSets
30.     Rule = {Rule, Neg(tr) | Neg(tr).supp >minsup, Neg(tr).conf > minconf };
31.   endif
32. endfor.

Fig. 2 Mining Negative Association Rules

The part I of the algorithm given in Figure 2 is capable of discovering all possible set of frequent itemsets subject to a user specified minimum confidence.

The part II of the algorithm given in Figure 2 is capable of finding all positive and negative association rules from the frequent itemsets subject to a user specified minimum confidence very quickly. The function *genRule(FreqSetk)* generates all positive subject to a user specified minimum confidence. The function *genNegCand(r)* generates all negative itemsets for the given positive association rule, using the support value calculated using the formula given in equations (1) and (3). From the generated negative item sets, the negative association rules are generated subject to the confidence calculated using the formula given in equations (2) and (4).

## IV. RESULTS AND DISCUSSION

The introduced algorithm has been implemented in Java and tested on Linux platform. A comprehensive experiment on spatial gene expression data has been conducted to study the impact of normalization.

A few sample records from spatial gene expression data (EMAGE) are listed in Table I. Different support-confidence thresholds were tested. A few positive and negative rules are listed in Table I. They are generated under the support and confidence constraints of 2% to 9% and 30% and 60%, respectively. Note that rules in the right column are negative rules discovered with respect to positive rules in the left column. Table II shows the number of positive rules and negative rules vs. user-specified minimum support and minimum confidence. Thus, the algorithm can successfully generate negative rules and the number of negative rules discovered is reasonable. The number of negative rules tends to be related to the number of positive rules given in Table II. However, it is inversely proportional to the minimum support threshold. The reason is less and less high-support 1-item set survives with increases in support threshold, which reduces the number of candidate negative rules significantly.

**TABLE I.** Spatial gene expression data from EAMGE database

| Uniqid | Gene name | EMAGE :1024 | EMAGE: 111 | EMAGE: 114 | EMAGE: 117 |
|---|---|---|---|---|---|
| **EMAGE:1024** | Cer1 | 1 | 0 | 0 | 0 |
| **EMAGE:111** | T | 0 | 1 | 1 | 0 |
| **EMAGE:114** | Mesp1 | 0 | 1 | 1 | 1 |
| **EMAGE: 117** | Pou5f1 | 0 | 0 | 1 | 1 |

**TABLE II.** Positive and negative rules generated (min_support=3% and min_confidence = 30% and 60% )

| Positive rules | Confidence | Negative rules | Confidence |
|---|---|---|---|
| T→Mesp1 | 100% | Cer1→ ¬ T | 100% |
| T→Pou5f1 | 50% | Cer1→ ¬Mesp1 | 100% |
| Mesp1→Pou5f1 | 67% | Cer1→ ¬Pou5f1 | 100% |

Given the number of positive rules P, the complexity of the algorithm is O(P). In this algorithm the complexity does not depend on the number of transactions since it is assumed that the supports of item sets have been counted and stored for use in this as well as other mining applications. However if we are considering discovering positive rules, which is necessary in generating negative rules, the algorithm must browse all combinations of items. The complexity of discovering positive rules depends on not only the number of transactions, but also the sizes of attribute domains as well as the number of attributes. The overall complexity will be proportional to that of discovering positive rules.

The performance is also affected by the choice of minimum support. A lower minimum support produces more numerous item sets and, with the same confidence constraint more positive rules will be generated, which adds to computation expense. The trend in the number of negative and the number of positive rules with different minimum support are shown in Figure 3.

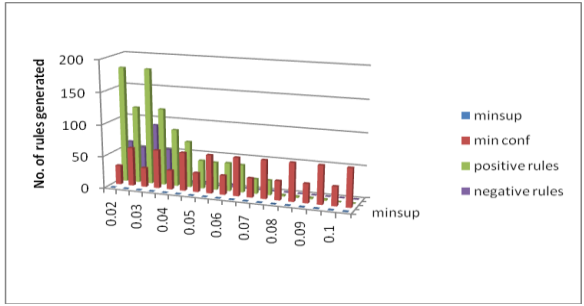

Figure 3. Positive and Negative Rules Discovered under Different Supports and Confidences

## V. CONCLUSION

In this paper, a novel method of mining positive and negative association rules from the spatial gene expression data has been introduced to generate frequently occur genes very quickly. The introduced algorithm does not produce candidate itemsets, it spends less time for calculating k-supports of the itemsets with the Boolean matrix pruned, and it scans the database only once and needs less memory space when compared with Apriori algorithm. The introduced algorithm is good enough for generating positive and negative association rules from spatial gene expression data very fast and memory efficient. Finally, the large and rapidly increasing compendium of data demands data mining approaches, particularly association rule mining ensures that genomic data mining will continue to be a necessary and highly productive field for the foreseeable future.

REFERENCES

[1] Baldock,R.A., Bard,J.B., Burger,A., Burton,N., Christiansen,J., Feng,G., Hill,B., Houghton,D., Kaufman,M., Rao,J. et al., "EMAP and EMAGE: a framework for understanding spatially organized data", Neuroinformatics, vol. 1, pp. 309–325, 2003.

[2] Pang-Ning Tan, Micahel Steinbach, Vipin Kumare, Intoduction to Data Mining Pearson Education, second edition, pp.74, 2008.

[3] Agrawal, R. & Srikant, R., "Fast Algorithms for Mining Association Rules in large databases". In Proceedings of the 20th International Conference on Very Large Databases pp. 487-499. Santiago, Chile, 1994.

[4] Agrawal, R., Imielinski, T., & Swami, A.,"Mining association rules between sets of items in large databases". Proceedings of the ACM SICMOD conference on management of data", Washington, D.C, 1993.

[5] Xu, Z. & Zhang, S., "An Optimization Algorithm Base on Apriori for Association Rules". Computer Engineering vol. 29(19), pp. 83-84, 2003.

[6] J. van Hemert and R. Baldock, "Mining spatial gene expression data for association rules", In S. Hochreiter and R. Wagner, editors, Proceedings of the 1st International Conference on BioInformatics Research and Development, Lecture Notes in Bioinformatics, pp.66–76. SpringerVerlag, 2007.

[7] Céline Becquet et al., "Strong-association-rule mining for large-scale gene-expression data analysis: a case study on human SAGE data", Genome Biology vol. 3(12):research0067.1–0067.16, November 2002.

[8] M.Anandhavalli, M.K.Ghose, K.Gauthaman, "Mining Spatial Gene Expression Data Using Association Rules", IJCSS vol. 3(5), 2009.

[9] S. Venkataraman, P. Stevenson, Y. Yang, L. Richardson, N. Burton, T. P. Perry, P. Smith, R. A. Baldock, D. R. Davidson, and J. H. Christiansen. Emage—edinburgh mouse atlas of gene expression: 2008 update. Nucleic Acids Research, 36(D):860–865, Jan 2008.

[10] J. van Hemert and R. Baldock. Matching spatial regions with combinations of interacting gene expression patterns. In M. Elloumi, J. K˙ung, M. Linial, R. Murphy, K. Schneider, and C. Toma, editors, Proceedings of the 2nd International Conference on BioInformatics Research and Development, volume 13 of Communications in Computer and Information Science, pages 347–361. Springer Verlag, 2008.